\documentstyle[12pt]{article}
\begin{document}

\begin{flushright}
ICRA-20-01-94 \\
\end{flushright}

\vspace{7mm}

\begin{center}
{\bf Analysis of Spots in  the {\it COBE} DMR
First Year Anisotropy  Maps}  \\
\vspace{3mm}
Sergio Torres\footnote{On leave from
Universidad de los Andes and Centro Internacional de F\'{i}sica,
Bogot\'{a}, Colombia.
E-mail: 40174::torres, torres@celest.lbl.gov} \\

\vspace{3mm}

{\footnotesize\it ICRA - International Center for
Relativistic Astrophysics,
Univiversit\`{a} di Roma, Piazzale Aldo Moro, 5 - 00185,
Rome, Italy} \\
\end{center}

\vspace{8mm}

\begin{center}
ABSTRACT
\end{center}

After the detection of structure in the microwave sky,
the characterization of the observed features may
be a useful guide to current and future experiments.
The first year {\it COBE} DMR sky maps were analyzed  in order
to identify the most significant hot and cold spots.
The statistical significance of potential spots of cosmic
origin was evaluated with
Monte Carlo simulations. The area, eccentricity, and location
of the most significant spots are given.

\begin{center}
{\sc 1. INTRODUCTION}
\end{center}

Anisotropies in the cosmic microwave background  (CMB) at large
angle scales have been detected by the Differential Microwave
Radiometers (DMR) on board the COBE
satellite~\cite{ben1,smoot1} and by the FIRS
experiment~\cite{ganga}.
Identifying and characterizing hot spots becomes now more relevant, and
experiments (such as TENERIFE) can design
survey strategies so as to look for a particular hot spot.
Knowing the geometry of hot spots may also be of great potential
for testing the `mixing of geodesics' effect~\cite{GU}.
However, due to the low signal-to-noise ratio on the DMR maps,
what appears as a hot spot is not necessarily
a legitimate hot  spot (i.e. one of cosmological origin or
one that can be attributed to a local source).

In an ideal experiment, with noiseless instruments,
a radiometer looking at the CMB fluctuations
should see hot regions which behave as unresolved sources.
It has been suggested that even noise limited experiments
may observe the hottest spots~\cite{sazhin}.
The expected number density of these hot
spots and their geometric
characteristics have been derived for 2D homogeneous Gaussian random
fields ~\cite{adler,bond,vit,col1,col2,col3,gott,martin}.

The excursion set of a random temperature field is the domain
of all points in which the field takes on values
$T \geq T_{\nu} = \nu \sigma$, where $\sigma$ and $\nu$
are the field standard deviation and threshold, respectively.
For high threshold levels $(\nu > 1)$, the excursion set is
characterized by non-connected regions (hot spots)
whose shape approaches a circle as the threshold increases.
Similarly, a cold spot can be defined as the region where
the temperature takes on values $T \leq -T_{\nu}$. A negative
$\sigma$  is used to indicate these cold spots.

\begin{center}
2. ANALYSIS
\end{center}

Data from the 53 and 90 GHz DMR radiometers,
the two with the best sensitivity, were used. The maps were
scaled to thermodynamic temperature,
the dipole was fitted and removed, the $2 \mu$K  kinematic quadrupole
was removed, and finally the maps were Gaussian smoothed
($\theta_{s} =  2.9^{\circ}$). A comparison of the fitted
dipole and quadrupole with the results of Smoot et al.~\cite{smoot1}
provided a test  for the integrity of the data and the
analysis software.  The  contribution of the galaxy is taken into account
by excluding  the equatorial band $|b| \leq 20^{\circ}$,
where $b$ is galactic latitude. After galactic cut  65\% of the  sky
remains available for analysis.
Sum (A+B) and difference (A--B)
maps are the sum  and difference
of the two  DMR maps at  each frequency.
Galactic and cosmic signals cancel out
in the (A--B) maps.
The mean temperature is subtracted after galactic cut.
The resulting temperature standard deviation of the 53 and 90 sum maps
are $51$ and $69 \, \mu$K respectively.
Only spots that appear at a threshold $|\nu| \geq 2.3$ were studied.

 Hot and cold spots on DMR maps were analyzed
by an algorithm that relies on the formation of tree data structures
on binary maps~\cite{torres1,torres2}. The statistical significance
of a spot is estimated by means of
Monte Carlo simulations that take into account instrument noise,
sky coverage, pixelization scheme and the DMR beam
characteristics~\cite{torres2}.
The location of a hot spot is  given by the coordinates
of the barycenter of the spot. The `eccentricity' parameter,
$\epsilon$,
is defined as the ratio of the distances from the center
of the spot to the
closest and furthest points along its contour.
This parameter coincides with the eccentricity of elliptical
patterns for large thresholds, but  has no straightforward
interpretation at low thresholds where contours are highly convoluted.
Shape information for
spots smaller or equal than a pixel is lost, and
a value of 1.0
is automatically assigned to its area and eccentricity.
Areas are given in pixel units, which for DMR are
$4\pi/6144 \approx 2.6^{\circ} \times 2.6^{\circ}$
per pixel.

Simulated sky maps of the microwave sky were generated
using a spherical harmonic expansion for the sky temperature,
$\Delta T/T = \Sigma_{\ell} \Sigma_{m} a_{\ell,m} W_{\ell} Y_{\ell,m}$,
with Gaussian random coefficients $a_{\ell,m}$ of zero mean and
model dependent variance. The weights $W_{\ell}$ for DMR
given by Wright et al.~\cite{wri1} were used.
Noise, determined by  instrument sensitivity and
the number of observations per pixel, is included in the simulations.

With a  power spectrum of primordial
density fluctuations $P(k) \propto k^n$, the variances of $a_{\ell,m}$
can be determined~\cite{bond}. Figure 1 shows the number of spots
for the DMR 53 (A+B) data and Monte Carlo results for noise
and for a $n=1$ model normalized to a quadrupole of $16 \, \mu K$.
{}From Fig. 1 it is clear that the number of spots descriptor
of the DMR data is in  good agreement with the presence of
cosmic signal. Some of the spots on DMR maps must therefore be cosmic.
Even though it is not possible to say whether a  spot is real or not,
one can assign a statistical significance to each spot  using
its height information in combination with the number of
standard deviations that its area deviates from the expected area
of  spots on noise maps.
The spot mean area and  variance in noise maps are
found with Monte Carlo simulations.

Due to noise alone, hot spots  can appear on DMR maps. However,
by the superposition of two or three maps, the probabilities
of finding the same `noise spot' in all maps is
substantially reduced.
Let $a_{\nu,i}$ denote the total area of the excursion set
of map $i$ at threshold level $\nu$.  The intersection
area of the excursion set of two maps,
$A_{\nu,ij} = a_{\nu,i} \cap a_{\nu,j}$,
is a quantity  that can be used to indicate the
presence of cosmic spots on a map.

Monte Carlo simulations
of skies as seen by DMR, if no cosmic signal was present,
were used to estimate $A_{\nu,ij}$ for noise alone.
Figure 2
shows the intersection area with threshold for the superposition
of the 53 and 90 sum and difference maps and the Monte Carlo data
for  simulations of noise. For large  thresholds $(> 1.5 \sigma)$
the statistics is low, but for small $\nu$ there is a significant
and systematic deviation  of the (A+B) data from what would be
expected for noise.
There are small differences among the Monte Carlo
data and the (A--B) points in Fig. 2. These can be attributed to the
sensitivity of $A_{\nu,ij}$ to the  level of smoothing of the
maps, which depends on  beam dilution
(not taken into account in the Monte Carlo simulations)
and  the inaccurate  estimation of beam smearing (due to
spacecraft drift during the $1/2$ second observation time).
If the (A--B) points of Fig. 2 are used as an estimate
of the expected $A_{\nu,ij}$ for noise, and in addition the Monte
Carlo error bars are used, one finds that the
excursion set intersection
area of the 53 and 90 (A+B) maps is 4.7, 3.1, and 1.7
standard deviations higher than expected for noise at
threshold levels 1.5, 1.75 and 2.0  respectively.

The area, location and eccentricity of the most
significant hot and cold spots are in Tables I and II.
The spots  are ordered according to
a `significance parameter' defined as
$S = \sum_{\nu}  |\nu| s_{\nu}$, where
$s_{\nu}$ is the number of standard deviations the spot area
deviates from the mean spot area of noise maps at threshold $\nu$.
If $s_{\nu} < 1.0$ it is set to 1.0 when computing $S$.
No absolute meaning should be given to $S$, it is only an
assessment of the relative probability that a spot may be of
cosmic origin.

At the $2.3 \sigma$ level there are no spots seen simultaneously
on the three (A+B) DMR maps. Only one hot spot (No. 4) and
one cold spot (No. 4) consistently appear on both 53 and 90 (A+B)
maps.
Hot spot No. 1 in 53 (A+B) map is clearly
associated with the galactic bulge at $l=0^{\circ}$ seen in this
map. The only features on the  31 GHz sum  map that may be
significant are a hot spot coincident with hot spot No. 6. and
one cold spot coincident with cold spot No. 10.
With the exception of hot spot No. 2 and the two spots
that appear in two maps (Hot No. 4 and Cold No. 4) at high $\nu$,
all other features are marginaly significant. Eliminating
hot spot No. 1 (galaxy) and all spots below $\nu=2.6$ there
remains 9 hot spots and 11 cold spots, consistent with
the expected results from simulations of cosmic signal
($n=1$, $16 \, \mu$K  Quadrupole).
With the four year DMR data  the signal-to-noise ratio
will be greater than one, thus cosmography will become  even more relevant,
and it will be possible to confirm or discard
the identified spots.  Bennet et al.~\cite{ben2} have shown
that the features on the first year maps are not correlated
with known astronomical  sources, thus if the spots
are real, their most likely origin is cosmic.

{\it Acknowledgments:} I thank  Prof. R. Ruffini for his
hospitality at ICRA. This work has been supported by
Colciencias of Colombia, project No. 1204-05-007-90 and
the European Community under contract No. CI1-CT92-0013.
The COBE datasets were developed by the NASA
Goddard Space Flight Center under the guidance of the COBE Science
Working Group and were provided by the NSSDC.

\newpage

\begin{flushleft}
{\Large\bf Figure Captions} \\
\vspace{5mm}
FIGURE 1. Number of spots on the $4\pi$ sphere for Monte Carlo noise
({\it squares}, upper curve),
a $n=1$ model ({\it squares}), and
the DMR (A+B) maps ({\it crosses})
scaled to the $4\pi$ sphere. \\

\vspace{3mm}

FIGURE 2. a) Excursion set intersection area for Monte Carlo noise
({\it squares}), 53 and 90 DMR sum maps ({\it dots}) and
difference maps ({\it crosses}), b) same as a) with different scale.
Area is in pixel units.

\end{flushleft}

\newpage

\begin{center}
TABLE I. Most significant hot spots \\
\begin{tabular}{rrrrrrrrr} \hline

N & S~~ & M & $\nu$~~ & A & $s_{\nu}~$  & $\epsilon$~~ & $l$~~ & $b$~~ \\
\hline
 1 & 141.5 & 53  & 2.33  & 26  & 29.1 & 0.33  &     4.4 &   25.5 \\
   &       & 53  & 2.67  & 16  & 15.3 & 0.36  &     4.8 &   24.8 \\
   &       & 53  & 3.00  & 12  &  8.6 & 0.26  &     4.0 &   24.4 \\
   &       & 53  & 3.33  &  2  &  0.1 & 0.50  &     2.6 &   22.5 \\
   &       & 53  & 3.67  &  1  & -0.6 & 1.00  &     1.3 &   22.6 \\
 2 &  45.4 & 53  & 2.33  & 12  & 11.2 & 0.60  &    51.9 &   65.7 \\
   &       & 53  & 2.67  &  8  &  6.1 & 0.70  &    54.2 &   65.6 \\
   &       & 53  & 3.00  &  2  & -0.3 & 0.50  &    47.3 &   65.6 \\
 3 &  15.2 & 53  & 2.33  &  7  &  4.8 & 0.37  &   184.4 &  -52.3 \\
   &       & 53  & 2.67  &  4  &  1.5 & 0.56  &   182.6 &  -52.1 \\
 4 &  14.5 & 53  & 2.33  &  6  &  3.5 & 0.74  &    50.3 &   38.6 \\
   &       & 53  & 2.67  &  4  &  1.5 & 0.54  &    51.1 &   38.6 \\
   &       & 90  & 2.33  &  4  &  0.9 & 0.72  &    46.4 &   38.9 \\
 5 &   8.0 & 53  & 2.33  &  1  & -2.9 & 1.00  &   160.1 &  -21.3 \\
   &       & 53  & 2.67  &  1  & -2.0 & 1.00  &   160.1 &  -21.3 \\
   &       & 53  & 3.00  &  1  & -1.2 & 1.00  &   160.1 &  -21.3 \\
 6 &   8.0 & 53  & 2.33  &  2  & -1.6 & 0.50  &   206.8 &  -20.5 \\
   &       & 53  & 2.67  &  2  & -0.8 & 0.50  &   206.8 &  -20.5 \\
   &       & 53  & 3.00  &  2  & -0.3 & 0.50  &   206.8 &  -20.5 \\
 7 &   5.0 & 53  & 2.33  &  4  &  1.0 & 1.00  &   194.6 &   44.1 \\
   &       & 53  & 2.67  &  1  & -2.0 & 1.00  &   195.6 &   42.3 \\
 8 &   5.0 & 53  & 2.33  &  2  & -1.6 & 0.50  &   212.6 &  -22.1 \\
   &       & 53  & 2.67  &  1  & -2.0 & 1.00  &   214.1 &  -21.9 \\
 9 &   5.0 & 53  & 2.33  &  3  & -0.3 & 0.37  &   258.2 &  -22.1 \\
   &       & 53  & 2.67  &  1  & -2.0 & 1.00  &   258.2 &  -22.1 \\
10 &   5.0 & 90  & 2.33  &  2  & -1.6 & 0.50  &     7.9 &  -25.1 \\
   &       & 90  & 2.67  &  2  & -0.8 & 0.50  &     7.9 &  -25.1 \\
11 &   2.3 & 53  & 2.33  &  1  & -2.9 & 1.00  &   340.3 &   51.3 \\
12 &   2.3 & 53  & 2.33  &  1  & -2.9 & 1.00  &   208.6 &   30.7 \\
13 &   2.3 & 53  & 2.33  &  1  & -2.9 & 1.00  &   310.5 &   35.7 \\
14 &   2.3 & 53  & 2.33  &  1  & -2.9 & 1.00  &   175.0 &  -49.6 \\
15 &   2.3 & 53  & 2.33  &  1  & -2.9 & 1.00  &    14.8 &  -48.8 \\
16 &   2.3 & 90  & 2.33  &  2  & -1.6 & 0.50  &   248.6 &   51.0 \\
17 &   2.3 & 90  & 2.33  &  4  &  0.9 & 0.95  &   164.1 &   23.1 \\
18 &   2.3 & 90  & 2.33  &  1  & -2.9 & 1.00  &   194.9 &   33.1 \\
19 &   2.3 & 90  & 2.33  &  1  & -2.9 & 1.00  &   175.5 &  -46.5 \\
20 &   2.3 & 90  & 2.33  &  2  & -1.6 & 0.50  &   160.7 &  -60.4 \\ \hline
\end{tabular} \\
\end{center}
\vspace{5mm}

{\footnotesize
NOTE - Hot spot characteristics as appear at different
threshold levels ($ \geq 2.3 $).
The `significance' $S$ and $s_{\nu}$ parameter are defined in the text.
The DMR map is indicated by M.
The area, A,  is in DMR pixel units.
The spot eccentricity is $\epsilon$, and
$l$,  $b$  are the galactic longitude and latitude in degrees.
The estimated error in pixel location is $\pm 1.5^{\circ}$.}

\newpage

\begin{center}
TABLE II. Most significant cold spots \\
\begin{tabular}{rrrrrrrrr} \hline

N & S~~ & M & $\nu$~~ & A & $s_{\nu}~$  & $\epsilon$~~ & $l$~~ & $b$~~ \\
\hline

 1 &  45.2 &  53 & -3.33 &   1 & -0.8 & 1.00  &   242.6 &   46.4 \\
   &       &  53 & -3.00 &   5 &  2.1 & 0.64  &   242.9 &   45.2 \\
   &       &  53 & -2.67 &   8 &  5.9 & 0.50  &   240.7 &   44.3 \\
   &       &  53 & -2.33 &  10 &  8.5 & 0.45  &   241.4 &   44.8 \\
 2 &  32.6 &  90 & -3.00 &   4 &  1.3 & 0.84  &   236.5 &   36.8 \\
   &       &  90 & -2.67 &   5 &  2.1 & 0.60  &   237.3 &   37.3 \\
   &       &  90 & -2.33 &  11 &  9.9 & 0.64  &   238.0 &   36.0 \\
 3 &  23.1 &  90 & -3.00 &   5 &  2.1 & 0.58  &   126.1 &  -51.3 \\
   &       &  90 & -2.67 &   5 &  2.1 & 0.58  &   126.1 &  -51.3 \\
   &       &  90 & -2.33 &   7 &  4.8 & 0.46  &   126.8 &  -51.1 \\
 4 &  16.7 &  53 & -2.33 &   1 & -2.8 & 1.00  &   256.3 &   64.0 \\
   &       &  90 & -2.67 &   4 &  1.2 & 0.97  &   247.6 &   70.1 \\
   &       &  90 & -2.33 &   7 &  4.8 & 0.44  &   250.3 &   68.2 \\
 5 &  14.7 &  53 & -2.67 &   4 &  1.4 & 0.51  &   336.0 &  -20.9 \\
   &       &  53 & -2.33 &   7 &  4.7 & 0.44  &   336.5 &  -22.1 \\
 6 &  13.6 &  53 & -2.67 &   1 & -1.9 & 1.00  &   280.0 &   55.5 \\
   &       &  53 & -2.33 &   7 &  4.7 & 0.42  &   281.2 &   55.8 \\
 7 &  13.6 &  53 & -2.67 &   1 & -1.9 & 1.00  &   275.1 &   75.5 \\
   &       &  53 & -2.33 &   7 &  4.7 & 0.52  &   277.3 &   74.3 \\
 8 &   5.1 &  90 & -2.33 &   5 &  2.2 & 0.42  &   100.9 &  -67.6 \\
 9 &   5.0 &  53 & -2.67 &   2 & -0.8 & 0.50  &   302.3 &   41.8 \\
   &       &  53 & -2.33 &   3 & -0.3 & 0.83  &   302.7 &   42.7 \\
10 &   5.0 &  90 & -2.33 &   4 &  1.0 & 0.50  &   242.4 &   22.9 \\
   &       &  90 & -2.67 &   2 & -0.8 & 0.50  &   243.2 &   23.0 \\
11 &   5.0 &  90 & -2.67 &   1 & -1.7 & 1.00  &   254.6 &   39.2 \\
   &       &  90 & -2.33 &   2 & -1.6 & 0.50  &  253.3 &   39.0 \\
12 &   2.3 &  53 & -2.33 &   2 & -1.6 & 0.50  &   288.4 &   66.4 \\
13 &   2.3 &  53 & -2.33 &   1 & -2.8 & 1.00  &    70.0 &  -26.7 \\
14 &   2.3 &  53 & -2.33 &   1 & -2.8 & 1.00  &    72.8 &  -24.3 \\
15 &   2.3 &  53 & -2.33 &   1 & -2.8 & 1.00  &   146.3 &  -41.3 \\
16 &   2.3 &  53 & -2.33 &   1 & -2.8 & 1.00  &   351.0 &  -52.5 \\
17 &   2.3 &  53 & -2.33 &   2 & -1.6 & 0.50  &   87.0 &  -46.5 \\
18 &   2.3 &  90 & -2.33 &   1 & -2.8 & 1.00  &   301.4 &   54.8 \\
19 &   2.3 &  90 & -2.33 &   2 & -1.6 & 0.50  &   270.0 &   61.8 \\
20 &   2.3 &  90 & -2.33 &   1 & -2.8 & 1.00  &   345.4 &  -27.4 \\
21 &   2.3 &  90 & -2.33 &   1 & -2.8 & 1.00  &   358.7 &   28.2 \\
22 &   2.3 &  90 & -2.33 &   1 & -2.8 & 1.00  &    88.6 &  -40.2 \\
23 &   2.3 &  90 & -2.33 &   1 & -2.8 & 1.00  &   197.3 &  -27.1 \\
24 &   2.3 &  90 & -2.33 &   1 & -2.8 & 1.00  &    84.0 &  -55.8 \\  \hline

\end{tabular} \\
\end{center}
\vspace{5mm}
{\footnotesize
NOTE - See note of Table I.}


\newpage

\begin{thebibliography} {99}
\bibitem{ben1} Bennett, C., et al., COBE-Preprint 94-01 (1994) and
astro-ph/9401012
\bibitem{smoot1} Smoot, G.F. et al, ApJ, 396, L1 (1992)
\bibitem{ganga} Ganga, K., Cheng, E., Meyer, S. \& Page, L.,
ApJ, 410, L57 (1993)
\bibitem{sazhin} Sazhin, M. V., MNRAS,  216, 25p (1985)
\bibitem{adler} Adler, R. J., The geometry of Random Fields
  (New York: Wiley) (1981)
\bibitem{bond} Bond, J. R., \& Efstathiou, G., MNRAS, 226, 655 (1987)
\bibitem{vit} Vittorio, N. \& Juszkiewicz, R.,  ApJ, 314, L29 (1987)
\bibitem{col1} Coles, P. \& Barrow, D.,  MNRAS, 228, 407 (1987)
\bibitem{col2} Coles, P., MNRAS, 231, 125 (1988)
\bibitem{col3} Coles, P., MNRAS, 234, 509 (1988)
\bibitem{gott} Gott, J. R., et al., ApJ, 352, 1 (1990)
\bibitem{martin} Mart\'{i}nez-Gonz\'{a}lez, E., \& Sanz, J. L.,
MNRAS, 237, 939 (1989)
\bibitem{GU} Gurzadyan, V. G., Kocharyan, A. A., Europhys. Lett., 22, 231
(1993)
\bibitem{torres1} Torres, S., ApJ Letters
({\it in press}) (1994) and astro-ph/9311067
\bibitem{torres2} Torres, S., in  The CMB Workshop, Edts. N. Mandolesi, et al.,
Capri, ({\it in press}) (1993)
\bibitem{wri1} Wright, E. L.,  et al., ApJ, 420, 1 (1994)
\bibitem{ben2} Bennett, C. L., et al., ApJ, 414, L77 (1993)
\end{thebibliography}
\end{document}